\documentclass[%
 reprint,
superscriptaddress,
 amsmath,amssymb,
 aps,
]{revtex4-2}
\usepackage{graphicx}
\usepackage{dcolumn}
\usepackage{color}
\usepackage{comment}
\usepackage{multirow}
\usepackage{hyperref}
\usepackage{float}
\makeatletter
\let\newfloat\newfloat@ltx
\makeatother
\usepackage{algorithm}
\usepackage{algpseudocode}
\usepackage{bm}
\usepackage{mathrsfs}
\usepackage{float}
\usepackage{amsmath}
\usepackage{warpcol}
\usepackage{longtable}
\graphicspath{{Images/}}
\usepackage{xcolor}
\definecolor{red1}{rgb}{0.6,0,0}

\usepackage{longtable,booktabs}

\begin{document}

\title{Uncovering multi-channel magnetic hopfion annihilation via a single-node, billion-spin-scale atomistic framework}

\newcommand{\KTH}{Department of Applied Physics, School of Engineering Sciences, KTH Royal Institute of Technology, 
AlbaNova University Center, SE-10691 Stockholm, Sweden}

\newcommand{\SeRC}{Swedish e-Science Research Center, KTH Royal Institute of Technology, SE-10044 Stockholm, Sweden}

\newcommand{\WISEKTH}{Wallenberg Initiative Materials Science for Sustainability (WISE), KTH Royal Institute of Technology, SE-10044 Stockholm, Sweden}

\author{Qichen Xu*}
    \affiliation{\KTH}
    \affiliation{\SeRC}
    \thanks{qichenx@kth.se}
\author{Anna Delin}
    \affiliation{\KTH}
    \affiliation{\SeRC}
    \affiliation{\WISEKTH}
\date{\today}

\begin{abstract}
Modern atomistic spin simulations combine long stochastic trajectories, thermodynamic sampling, static optimization and multi-image transition-path workflows, all of which rely on repeated evaluation of spin Hamiltonians and become computationally prohibitive on the large lattices required for three-dimensional magnetic textures. We introduce SpinX, a GPU-native atomistic spin simulation framework built around a unified Hamiltonian interface and multiple user-selectable computational backends. Its core is a crystallographic sublattice decomposition that reformulates translationally invariant spin interactions as multi-channel tensor convolutions, enabling dense, sparse and FFT-based convolution backends, while irregular systems are handled by pair-list evaluation and long-range dipolar fields by reciprocal-space FFT. Implemented in JAX, SpinX supports deterministic and stochastic Landau--Lifshitz--Gilbert dynamics, Monte Carlo sampling, static optimization, dynamical spectroscopy and string and geodesic nudged elastic band transition-path calculations on heterogeneous accelerator platforms. A validated mixed-precision mode combines fp32 field evaluation with fp64 spin-state propagation. We validate SpinX against analytical single-spin dynamics, finite-size thermodynamics of bcc Fe and transverse dynamic structure factors. Performance benchmarks show peak throughput exceeding 10 billion spin-site operations per second on a single accelerator and aggregate single-node workloads of over 1 billion atomic spins. Applying this framework to an exchange-stabilized magnetic hopfion, we uncover two competing annihilation channels on a million-spin atomistic lattice: a previously reported axial-collapse pathway and a distinct lateral-rupture pathway with a different transition morphology and activation barrier. By combining accelerator-native throughput with large-scale transition-state workflows, SpinX establishes atomistic spin simulation as a practical route to studying three-dimensional magnetic textures and their energy landscapes.
\end{abstract}
\maketitle

\begin{figure*}
    \centering
    \includegraphics[width=16cm]{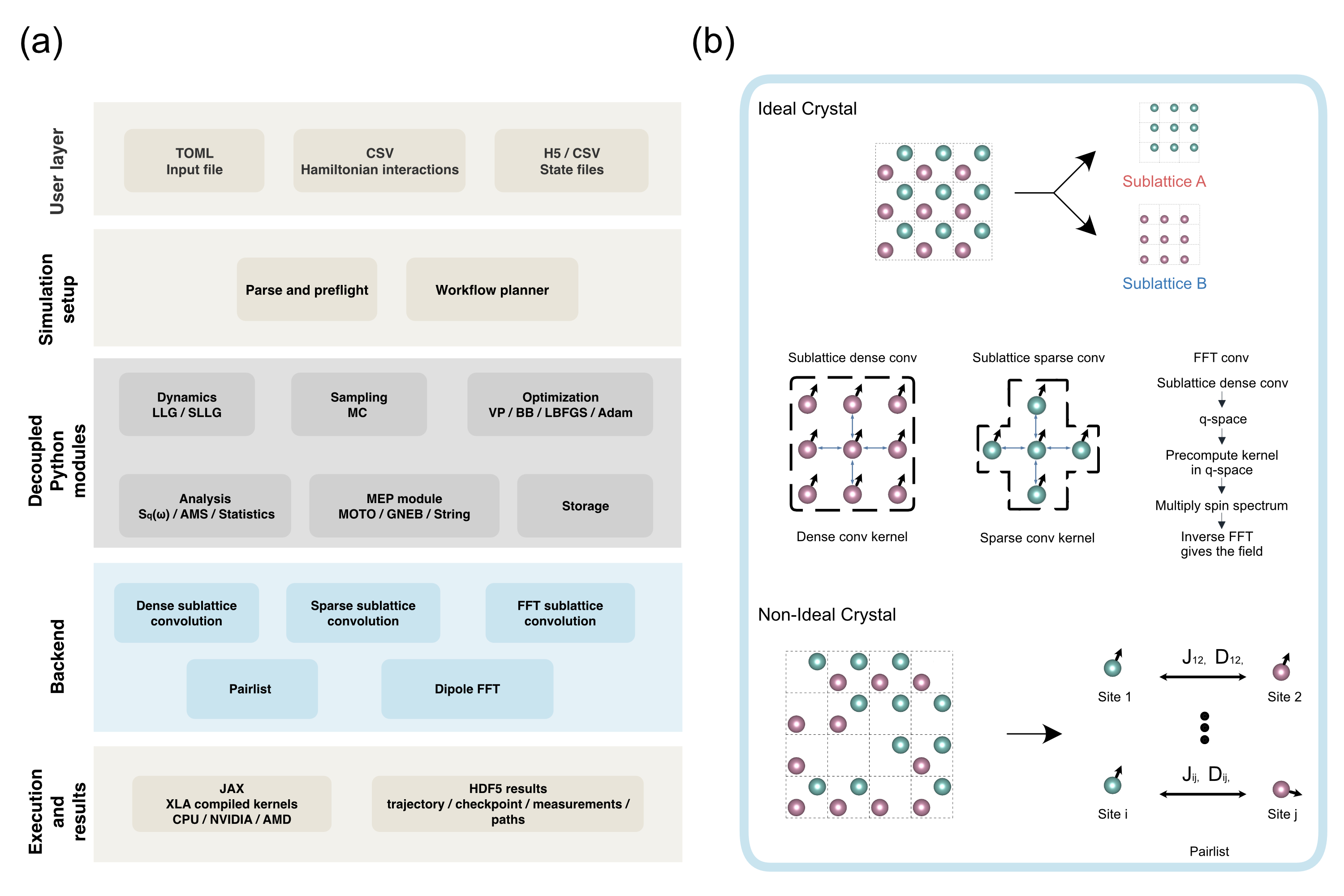}
    \caption{Modular architecture and backend dispatch in SpinX.
(a) Overview of the SpinX software stack. User inputs, including TOML configuration files, tabulated Hamiltonian parameters, and spin-state files, are parsed and checked in the simulation setup layer. The physics functionality is organized as decoupled Python modules for spin dynamics, thermodynamic sampling, static optimization, analysis, minimum-energy-path calculations, and data storage. These modules call a unified Hamiltonian backend, which dispatches effective-field evaluations to specialized kernels, including sublattice dense convolution, sublattice sparse convolution, FFT-based convolution, pairlist evaluation, and dipolar FFT. JAX/XLA compilation provides hardware-portable execution on CPUs, NVIDIA GPUs, and AMD GPUs, while trajectories, checkpoints, path data, and computed observables are stored in HDF5 format.
(b) Backend selection for effective-field evaluation. For ideal crystals with translational symmetry, SpinX decomposes the magnetic lattice into sublattice channels (analogous to colour channels in image processing) and maps pairwise bilinear interactions onto dense, sparse, or FFT-based tensor convolutions. The FFT backend precomputes the interaction kernel in reciprocal space and evaluates the field by multiplying the spin spectrum followed by an inverse FFT. Long-range dipolar interactions use the same reciprocal-space convolution principle through a dedicated dipole-FFT backend. For non-ideal crystals, including chemical disorder, structural defects, and explicit site-to-site connectivity, SpinX uses a generalized pairlist representation that stores local Hamiltonian parameters such as the exchange \(J_{ij}\) and Dzyaloshinskii--Moriya vectors \(\mathbf{D}_{ij}\).}
\label{fig:spinx-overview}
\end{figure*}

\begin{figure*}
    \centering
    \includegraphics[width=16cm]{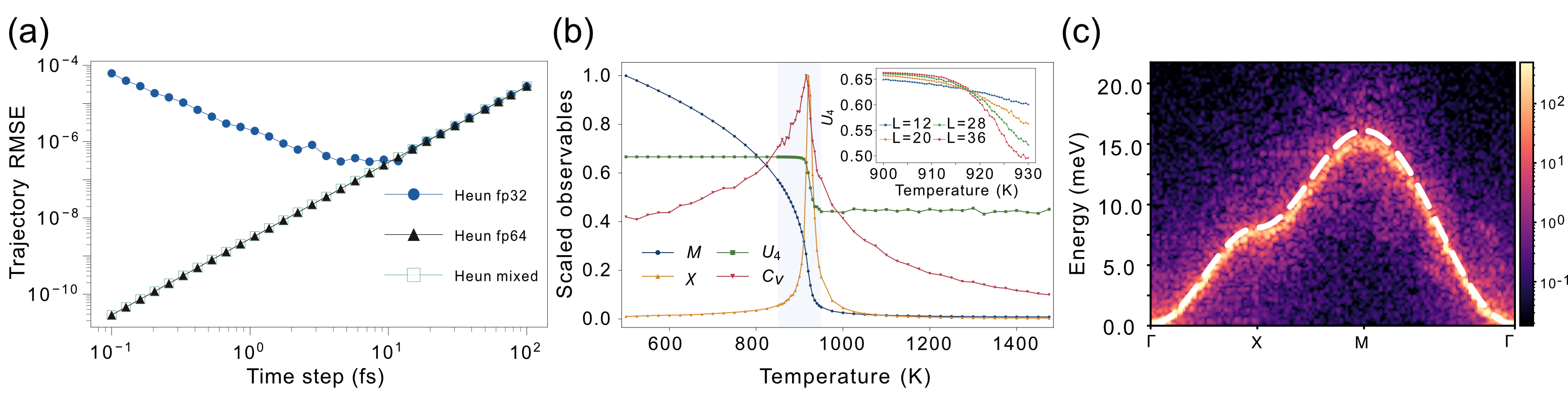}
    \caption{Numerical validation of SpinX using analytical, thermodynamic, and dynamical benchmarks.
(a) Single-spin deterministic LLG benchmark. The trajectory root-mean-square error is computed over a 500 ps trajectory sampled every 1 ps by comparing Heun integration with the analytical precession solution for time steps \(\Delta t = 0.1\)-\(100\,\mathrm{fs}\). In mixed-precision mode, effective-field evaluations are performed in single precision, while the spin state is stored and advanced in double precision. This mixed-precision scheme closely follows the fp64 baseline over the tested range, whereas pure fp32 exhibits a round-off error floor at small time steps. (b) Thermodynamic validation by Monte Carlo sampling for a classical bcc Fe Heisenberg model. The magnetization \(M\), Binder cumulant \(U_4\), magnetic susceptibility \(\chi\), and specific heat \(C_v\) are shown as functions of temperature, \(M\), \(\chi\), and \(C_v\) are scaled for visual comparison. The shaded region marks the estimated critical-temperature region. The inset shows Binder-cumulant crossings for system sizes \(L = 12\)-\(36\), used for finite-size estimation of the ferromagnetic-to-paramagnetic transition temperature. (c) Dynamical validation from the transverse dynamic structure factor. The spectrum \(S(q,\omega)=S_{xx}(q,\omega)+S_{yy}(q,\omega)\) is evaluated along the high-symmetry \(\Gamma\)-\(X\)-\(M\)-\(\Gamma\) path, with intensity shown in arbitrary units on a logarithmic scale. The white dashed curve denotes the adiabatic magnon spectrum (AMS) calculated from linear spin-wave theory, which follows the dominant simulated spectral weight across the Brillouin-zone path.}
    \label{fig:validation}
\end{figure*}

\begin{figure*}
    \centering
    \includegraphics[width=16cm]{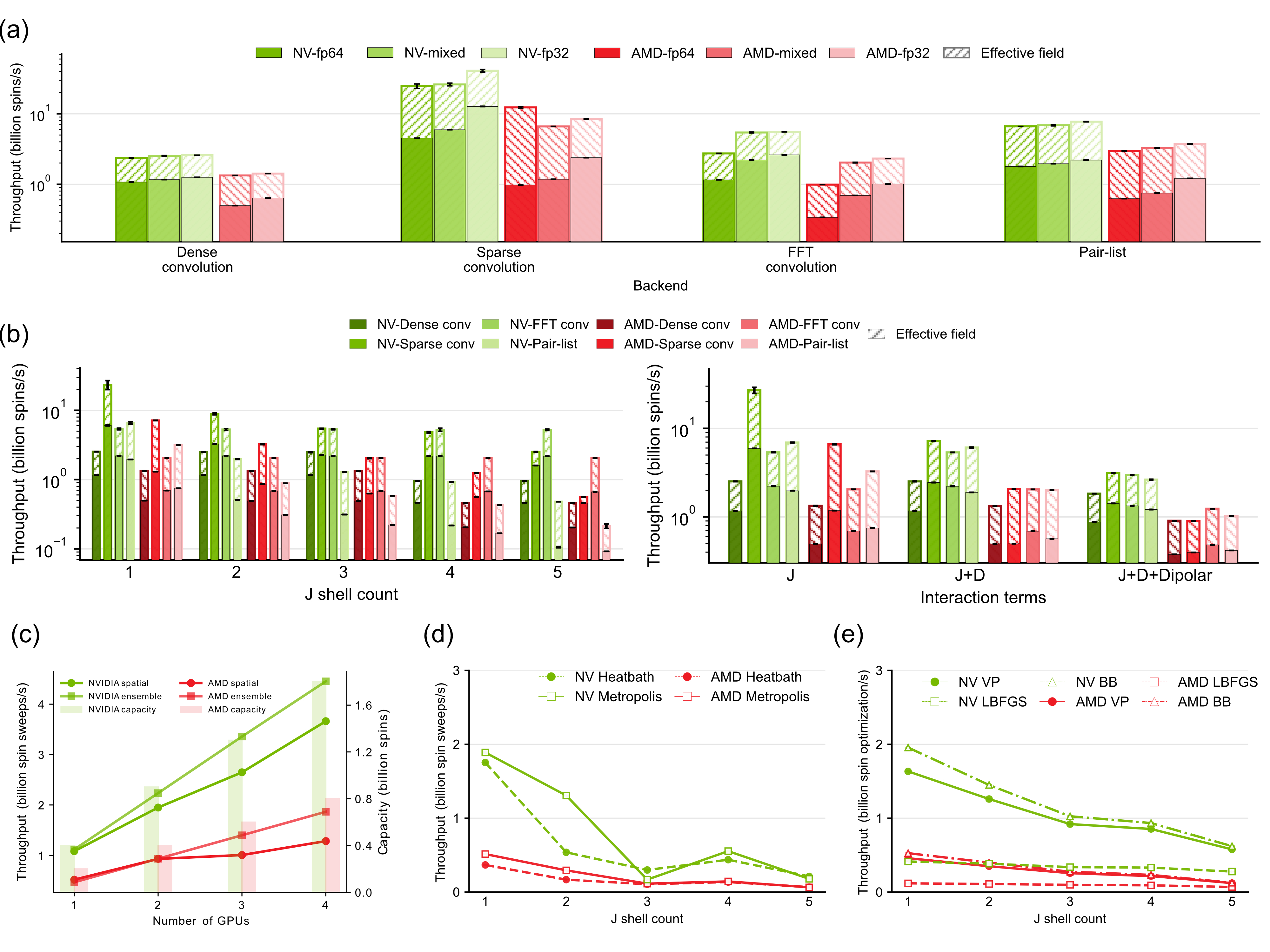}
    \caption{GPU performance and scaling benchmarks of SpinX. Benchmarks were performed on NVIDIA GH200 GPUs (Dardel, KTH, GPU portion only) and AMD MI250X accelerators (LUMI, CSC). For MI250X, throughput is reported per Graphics Compute Die (GCD), each physical card contains two GCDs. Unless otherwise stated, timings exclude JIT compilation and include device synchronization. Error bars denote the standard deviation over repeated executions.(a) Single-device backend throughput for a nearest-neighbour exchange model. Benchmarks are executed on a $200 \times 200 \times 200$ spin system. Dense convolution, sparse convolution, FFT convolution, and pairlist backends are compared on NVIDIA and AMD hardware in fp64, mixed-precision, and fp32 modes. Solid bars show Heun integration throughput, whereas hatched bars show isolated effective-field evaluation throughput, the two quantities should be interpreted independently rather than stacked. Missing bars indicate unsupported backend/precision combinations.(b) Dependence of throughput on Hamiltonian complexity. Left, mixed-precision throughput as the exchange-interaction shell count increases from one to five. Right, throughput for nearest-neighbour models containing exchange only ($J$), exchange plus Dzyaloshinskii–Moriya interaction ($J+D$), and exchange plus DMI plus long-range dipolar interactions ($J+D+\text{dipolar}$). Solid and hatched bars denote Heun integration and isolated effective-field evaluation, respectively.(c) Single-node multi-GPU scaling. Spatial domain decomposition is benchmarked using up to four devices for a $204 \times 200 \times 200$ nearest-neighbour model, with the first dimension chosen to be divisible by the number of devices. Ensemble scaling distributes independent trajectories across devices. Translucent background bars, read against the right axis, indicate the maximum number of spins that fit in memory for the mixed-precision dense-convolution workflow.(d) Monte Carlo sampling throughput as a function of exchange shell count. Heat-bath and Metropolis updates are compared on NVIDIA and AMD hardware, with throughput reported as billion attempted spin updates per second.(e) Static-minimization throughput as a function of exchange shell count. Velocity projection (VP), Barzilai–Borwein (BB), and L-BFGS are benchmarked over a fixed number of iterations, with throughput reported as billion spin-site updates per second.}
        \label{fig:benchmark}
    \end{figure*}

\begin{figure*}
    \centering
    \includegraphics[width=16cm]{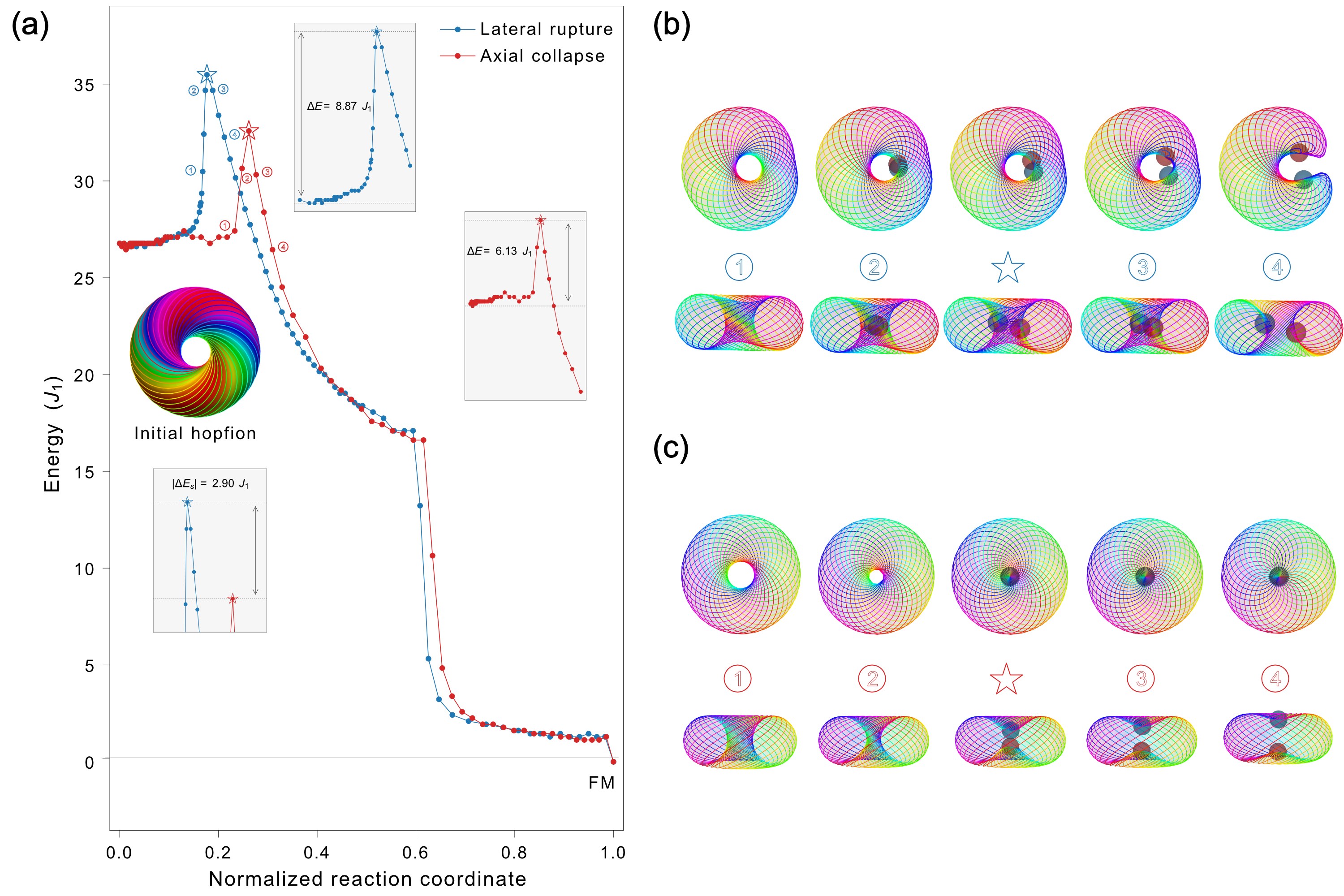}
    \caption{Two competing annihilation pathways of magnetic hopfions.
(a) Refined transition paths connecting the initial hopfion state to the ferromagnetic (FM) state. Both pathways are obtained by string-method initialization followed by GNEB refinement on an atomistic lattice, with open boundary conditions and the hopfion initialized near the center of the simulation box, the large surrounding ferromagnetic region reduces spurious boundary effects on the annihilation mechanism. The blue path corresponds to a lateral-rupture channel, whereas the red path corresponds to an axial-collapse channel. Reaction coordinates are normalized independently for each path. The lateral-rupture saddle region was further refined using a denser image distribution and stringent CI-GNEB relaxation to improve the resolution of the local rupture process. Energies are shown in units of the nearest-neighbour exchange constant \(J_1\). Open stars mark the highest-energy images along the two paths. Insets show magnified views of the barrier regions, with activation barriers measured relative to the local metastable minimum on the hopfion side. The lower inset highlights the saddle-region energy offset between the two channels.
(b) Representative configurations along the lateral-rupture path. The 3D spin textures are visualized using cross-sectional spin-phase tubes derived from the \(m_z=0\) isosurface, with color representing the local spin phase. The hopfion develops a side-localized rupture, where Bloch-point-like singular regions mediate the topological opening of the toroidal texture before relaxation toward the FM state. Numbered labels correspond to the sequential images marked in panel (a), with the star denoting the highest-energy image.
(c) Representative configurations along the axial-collapse path. The hopfion core shrinks along the axial direction and collapses toward the FM state through a more symmetric channel. The visualization style matches panel (b), and the numbered labels correspond to the red markers in panel (a).}
\label{fig:hopfion}
\end{figure*}

\section*{Main}

Many phenomena in modern magnetism --- including lattice-scale frustrated exchange, non-collinear spin textures, thermal fluctuations and topological singularities such as Bloch points --- couple atomic-scale interactions to mesoscale magnetic textures \cite{Antropov1995PRL,Antropov1996PRB,xu2025design}. Characterizing the stability and energy landscapes of such textures requires integrated simulation workflows that combine long stochastic trajectories, thermodynamic sampling, static minimization and multi-image transition-path searches \cite{Sallermann2023PRB,Lobanov2023PRB,xu2026general,xu2023metaheuristic,xu2023genetic}. This simultaneous demand for large spatial volumes, atomistic lattice resolution and repeated path evaluations defines a computational regime that is not well served neither by small atomistic cells nor by continuum descriptions that smooth over lattice-scale singular events \cite{Kuchkin2025PRR}. Three-dimensional topological solitons such as magnetic hopfions are a particularly demanding example \cite{Liu2018PRB,Wang2019PRL,Kent2021NatCommun,Zheng2023Nature}. Their stability is determined not only by the relaxed metastable texture, but also by the topology and energetics of the annihilation pathways connecting them to the trivial ferromagnetic state \cite{Sallermann2023PRB,Lobanov2023PRB}. Resolving such pathways in practice can require million-spin-scale lattices to separate the texture from spurious boundary effects while preserving atomistic resolution \cite{Sallermann2023PRB,Lobanov2023PRB}.

Several mature simulation packages have shaped the current landscape of atomistic and continuum spin modelling. UppASD and VAMPIRE provide established platforms for atomistic spin dynamics, thermodynamic sampling and finite-temperature magnetic phenomena, while Spirit integrates atomistic dynamics with Monte Carlo sampling, transition-path methods and visualization \cite{eriksson2017atomistic,xu2023spinview, evans2014atomistic, Mueller2019PRB}. On the continuum side, mumax3 has demonstrated the transformative impact of GPU acceleration for high-throughput micromagnetic simulations. These developments have made many classes of spin simulations routine. However, emerging workflows for three-dimensional magnetic textures increasingly require a different combination of capabilities, including atomistic lattice resolution, accelerator-native execution across heterogeneous GPU architectures, multiple mathematically equivalent Hamiltonian backends, a unified treatment of regular and irregular interaction graphs, automatic-differentiation compatibility and multi-image transition-path calculations at the million-spin-per-image scale. Existing atomistic codes typically rely on solver-specific data structures and explicit interaction lists, which are general but do not naturally expose regular crystal Hamiltonians as convolutional or FFT-convolutional workloads. A framework that combines these capabilities within a single compiler-driven Hamiltonian engine would make large-scale atomistic workflows more composable, portable and directly compatible with differentiable programming and machine-learning interfaces.

Here we introduce SpinX, a GPU-native atomistic spin simulation framework designed for high-throughput and large-scale calculations on modern accelerators. The central design principle of SpinX is to express spin Hamiltonians through a unified interface that covers scalar and tensorial bilinear interactions as well as higher-order terms such as biquadratic, scalar-chirality and quadruplet interactions, and to evaluate the corresponding effective fields through user-selectable computational backends. For ideal crystals with translational symmetry, crystallographic sublattice decomposition reformulates translationally invariant pair interactions as multi-channel tensor convolutions: each crystallographic sublattice plays the role of a colour channel in image processing, and interactions between sublattices are encoded as cross-channel kernels. This mapping enables dense, sparse and FFT-based convolution backends to be implemented using Accelerated Linear Algebra (XLA) convolution, FFT and fused array primitives, while preserving a common Hamiltonian interface\cite{50530}. For non-ideal crystals with chemical disorder, structural defects or explicit site-resolved couplings, the same Hamiltonian representation is evaluated through a generalized pair-list backend. Long-range dipolar interactions are treated by a reciprocal-space FFT convolution backend. This backend structure allows SpinX to exploit lattice regularity when available, while retaining a general representation for irregular lattice settings.

SpinX is implemented in JAX and compiled through XLA, enabling hardware-portable execution across heterogeneous accelerator platforms, including both NVIDIA and AMD GPUs\cite{jax2018github}. To our knowledge, SpinX is the first atomistic spin-dynamics framework to systematically employ a mixed-precision execution mode in which effective-field evaluation is performed in fp32 while spin-state propagation is retained in fp64. This separates the most bandwidth-intensive part of the calculation from the accuracy-sensitive integration state, reducing memory traffic and exploiting the lower-precision arithmetic that modern accelerators natively accelerate, while preserving long-time integration accuracy. The framework follows a modular functional design: deterministic and stochastic Landau--Lifshitz--Gilbert (LLG) dynamics, heat-bath and Metropolis Monte Carlo sampling, static optimization, spectral analysis and minimum-energy-path calculations all call the same shared Hamiltonian engine. This avoids duplicating physics logic across solvers and allows independent modules and backends to be cross-validated against one another. Because the framework is written in JAX, registered Hamiltonian terms compose naturally with automatic differentiation, batching and just-in-time compilation, providing a direct interface to gradient-based optimization, sensitivity analysis and machine-learning workflows.

We validate SpinX using analytical, thermodynamic and dynamical benchmarks. Single-spin deterministic dynamics are benchmarked against the analytical LLG precession solution across multiple time-step and precision settings \cite{Antropov1995PRL,Antropov1996PRB}. Thermodynamic sampling is validated through heat-bath Monte Carlo simulations of a classical body-centred cubic (bcc) Fe Heisenberg model, including a Binder cumulant finite-size scaling analysis of the ferromagnetic transition \cite{Pajda2001PRB}. Dynamical correlations are tested by computing the transverse dynamic structure factor and comparing the dominant simulated spectral weight with the adiabatic magnon spectrum obtained from linear spin-wave theory \cite{Pajda2001PRB}. Together, these tests verify both the physical accuracy of the solvers and the consistency of the shared Hamiltonian engine.

Performance benchmarks show that SpinX achieves peak throughput exceeding 10 billion spin-site operations per second on a single accelerator and supports aggregate single-node workloads of over 1 billion atomic spins. This scale is particularly vital for path-based calculations, where a multi-image path-ensemble multiplies the effective number of spin degrees of freedom. Rather than exposing a single universally fastest kernel, the framework provides a set of mathematically equivalent backends with distinct performance profiles. Dense convolution, sparse convolution, FFT convolution, pair-list evaluation and dipolar FFT each dominate in different regions of interaction range, lattice regularity, floating-point precision and memory footprint. This regime-based architecture is essential because different physical workloads--ranging from short-range dynamics and extended frustrated exchange to long-range dipolar interactions, thermal sampling and static minimization--each present unique computational bottlenecks.

As a demonstration of the scientific capability enabled by this framework, we investigate annihilation pathways of an exchange-stabilized magnetic hopfion in a minimal frustrated-exchange model \cite{Sallermann2023PRB,Lobanov2023PRB}. The goal is not to provide material-specific predictions, but to resolve competing topological collapse mechanisms in a controlled atomistic setting. Previous transition-state studies of exchange-stabilized hopfions have focused on an axial-collapse mechanism, in which Bloch-point-like defects nucleate near the hopfion axis and mediate collapse of the toroidal texture \cite{Sallermann2023PRB,Lobanov2023PRB}. Using string-method initialization followed by geodesic nudged elastic band refinement on a ($100 \times 100 \times 100$) atomistic lattice, corresponding to ($10^6$) spins per image, under open boundary conditions, we identify a second annihilation channel in the same Hamiltonian: a lateral-rupture pathway. In this pathway, the hopfion tube develops a side-localized constriction and is severed by localized Bloch-point-like singular regions away from the central axis before relaxing toward the ferromagnetic state. This process is distinct from boundary escape, as the rupture occurs on the hopfion tube rather than through the sample boundary \cite{Lobanov2023PRB}. To our knowledge, such a lateral-rupture mechanism has not been reported for exchange-stabilized hopfions in frustrated-exchange atomistic models. The existence of two competing transition states with distinct activation barriers expands the picture of hopfion decay and illustrates how integrated atomistic workflows at the million-spin-per-image scale can uncover topological transition channels that are difficult to resolve without a unified large-scale simulation workflow.

\subsection*{A unified Hamiltonian interface with multiple computational backends}

SpinX is organized around a shared Hamiltonian interface rather than solver-specific implementations. Energy, effective-field and torque evaluations form the common computational core used by deterministic and stochastic Landau--Lifshitz--Gilbert (LLG) dynamics, Monte Carlo sampling, static minimization, spectral analysis and minimum-energy-path calculations. As shown in Figure~1a, user-provided configurations, interaction tables and spin-state files are first processed by the simulation setup layer, which validates the model and constructs the requested workflow. The resulting spin Hamiltonian is then accessed by decoupled Python modules for dynamics, sampling, optimization, analysis and minimum-energy-path calculations, while storage modules record trajectories, checkpoints and observables. The computational modules differ in how they manipulate energies, fields, torques or trajectories, but they call the same shared Hamiltonian engine where physical model evaluations are required.

The backend layer provides distinct numerical realizations of the same physical model. SpinX supports scalar and tensorial bilinear interactions, together with higher-order terms such as scalar biquadratic, scalar-chirality and quadruplet interactions in supported workflows. For pairwise bilinear terms, a unified $3 \times 3$ interaction tensor $\mathcal{J}_{ij}^{\alpha\beta}$ provides a common representation for isotropic exchange, symmetric anisotropic exchange and antisymmetric Dzyaloshinskii--Moriya interactions. This tensor representation maps directly onto local parameters such as the exchange scalar $J_{ij}$ and the Dzyaloshinskii--Moriya vector $\mathbf{D}_{ij}$ (Figure~1b), allowing the same Hamiltonian to be evaluated by different backends without modifying the high-level physics modules that call it.

For ideal crystals with translational symmetry, SpinX uses crystallographic sublattice decomposition (Figure~1b). The magnetic lattice is decomposed into discrete sublattice channels, conceptually analogous to colour channels in image processing, and translationally invariant pair interactions are encoded as cross-channel convolution kernels. In this representation, effective-field evaluation becomes a multi-channel tensor-convolution problem. The dense convolution backend handles compact regular kernels. The sparse convolution backend exploits structured sparse-shell interaction patterns. The FFT convolution backend treats extended translationally invariant interactions in reciprocal space. For the FFT backend, the interaction kernel is precomputed in $q$-space, multiplied by the transformed spin spectrum and transformed back to real space to obtain the effective field. Long-range dipolar interactions use the same reciprocal-space convolution principle through a dedicated dipole-FFT path in supported dynamics workflows.

For non-ideal crystals, translational symmetry can be broken by chemical disorder, structural defects, vacancies or explicitly site-resolved couplings. In such heterogeneous settings, tensor convolution is no longer always the natural data structure. SpinX therefore provides a generalized pairlist backend, in which interactions are evaluated from explicitly indexed site-to-site couplings and their corresponding Hamiltonian entries. This representation preserves the same abstract Hamiltonian interface used by the convolutional backends while providing the structural flexibility required for site-dependent interactions and explicitly indexed interaction patterns.

The choice of computational backend is user-controlled. SpinX does not assume a single universally optimal kernel. Instead, it exposes several numerically consistent realizations of the same Hamiltonian core within their overlapping capability domains. Dense convolution, sparse convolution, FFT convolution, dipole FFT and pairlist evaluation are suited to different combinations of lattice geometry, interaction range, floating-point precision and hardware platform. The performance benchmarks in Figure~3 provide practical guidance for selecting a backend for a given simulation regime.

\subsection*{Validation across dynamics, thermodynamics and spectroscopy}

We next validate SpinX at the level of time integration, statistical sampling and dynamical correlations. These tests probe different components of the framework, including the deterministic LLG integrator, the mixed-precision execution mode, the Monte Carlo sampling routines and the spectral post-processing pipeline. Because all three benchmarks use the same model parsing layer and shared Hamiltonian engine, their agreement provides an end-to-end validation of the simulation stack.

We first consider the simplest deterministic setting with an analytical reference: a single spin precessing under a uniform, constant magnetic field. Figure~2a shows the trajectory root-mean-square error (RMSE) of the Heun integration scheme over a 500~ps trajectory sampled every 1~ps, plotted as a function of the integration time step \(\Delta t\). In fp64, the error follows the expected time-discretization trend as the time step is reduced. The mixed-precision mode, in which effective-field evaluations are performed in fp32 while the spin state is stored and advanced in fp64, closely follows the fp64 baseline over the tested range. In contrast, the pure fp32 calculation develops a round-off error floor at small time steps, where further reducing \(\Delta t\) no longer improves the trajectory error. This benchmark verifies the deterministic integrator and supports the use of mixed precision in large-scale production runs.

Thermodynamic sampling is validated using a classical bcc Fe Heisenberg model with a multi-shell exchange profile. Figure~2b shows the temperature dependence of the magnetization \(M\), Binder cumulant \(U_4\), magnetic susceptibility \(\chi\) and specific heat \(C_v\). The magnetization decreases across the transition region, while \(\chi\) and \(C_v\) show enhanced fluctuations in the same temperature window. The Binder-cumulant analysis provides a finite-size estimate of the ferromagnetic-to-paramagnetic transition: the inset shows the crossing behaviour of \(U_4\) for system sizes \(L=12,20,28\) and 36. This calculation tests the heat-bath Monte Carlo implementation, local energy and field updates, and the ability of SpinX to reproduce standard finite-temperature behaviour in an atomistic Heisenberg model.

We further evaluate the dynamical correlation pipeline by computing the transverse dynamic structure factor,
\(S(q,\omega)=S_{xx}(q,\omega)+S_{yy}(q,\omega)\), at \(T=5\,\mathrm{K}\) along the high-symmetry \(\Gamma\)--\(X\)--\(M\)--\(\Gamma\) path. 
The low temperature suppresses strong thermal broadening, while the transverse projection removes the static longitudinal magnetization component, making this benchmark suitable for direct comparison with the zero-temperature adiabatic magnon spectrum. As shown in Figure~2c, the dominant simulated spectral weight follows the adiabatic magnon spectrum (AMS) obtained from linear spin-wave theory. This agreement validates the generation of spin-dynamics trajectories, the space--time Fourier analysis and the reciprocal-space path construction used in the spectroscopy module. It also checks the sign conventions, coordinate transformations and energy-to-frequency conversion used to compare finite-temperature spin dynamics with the zero-temperature spin-wave reference.

Together, these analytical, thermodynamic and spectroscopic benchmarks validate the main physical solvers in SpinX and show that the shared Hamiltonian engine gives consistent results across deterministic dynamics, thermal sampling and dynamical response calculations. Backend-to-backend consistency of energies and effective fields across the dense, sparse, FFT and pairlist backends is documented in the Supplementary Information.

\subsection*{GPU performance regimes and scaling}

We then benchmark the computational performance of SpinX across backend choices, precision modes, hardware platforms and simulation workflows. All timings exclude JIT compilation and include device synchronization after execution. Throughput is reported as the number of spin sites processed per second, with the precise operation defined by each benchmark: field evaluations, Heun spin steps, Monte Carlo attempted spin updates or optimization iterations. For AMD MI250X, results are reported per Graphics Compute Die (GCD), whereas NVIDIA results use the GPU portion of the NVIDIA GH200 Grace Hopper Superchip.

Figure~3a compares the single-device throughput of the four main Hamiltonian backends on a \(200 \times 200 \times 200\) nearest-neighbour exchange model. The solid bars report full Heun integration throughput, while the hatched bars report isolated effective-field evaluation throughput. These two measurements should be interpreted independently. The field-only benchmark measures the raw throughput of the Hamiltonian backend, whereas the full Heun benchmark additionally includes the predictor-corrector update, torque evaluation and spin normalization. Several backend and precision combinations exceed \(10^9\) spin sites per second for this short-range model, with the sparse convolution backend giving the highest throughput in the nearest-neighbour regime. The effect of precision is backend and platform dependent, while mixed precision provides a practical trade-off between speed, memory use and the accuracy behaviour established in Figure~2a.

The relative performance of the backends changes as the Hamiltonian becomes more complex. Figure~3b, left, shows the mixed-precision throughput as the number of exchange shells is increased from one to five. This benchmark illustrates why SpinX exposes multiple backend realizations instead of a single default kernel. Backends designed for compact nearest-neighbour interactions are not necessarily optimal for extended frustrated-exchange models, and the preferred backend changes with interaction range and memory-access pattern. Figure~3b, right, compares models with exchange only, exchange plus Dzyaloshinskii--Moriya interaction, and exchange plus DMI plus long-range dipolar fields. Adding DMI increases the local tensor structure of the field evaluation, while the dipolar term introduces a reciprocal-space contribution. The resulting overhead depends on the backend because different implementations reuse intermediate data, memory layouts and Fourier-space representations differently.

We then evaluate single-node multi-GPU scaling for the mixed-precision dense-convolution workflow (Figure~3c). Two scaling modes are considered. In spatial domain decomposition, the lattice is partitioned across devices and each device advances a subdomain of the same simulation. In ensemble scaling, independent trajectories are distributed over devices. Ensemble scaling provides an upper-bound reference because it requires little inter-device communication, whereas spatial decomposition tests distributed execution of a single large simulation. Both modes scale across multiple devices on NVIDIA and AMD hardware. The translucent background bars (right axis) indicate the maximum system size that fits in memory, demonstrating that multi-GPU execution increases not only throughput but also the accessible spin capacity.

Monte Carlo sampling has a different performance profile from deterministic time integration, since each update touches only a single spin and its neighbours rather than the entire lattice. Figure~3d reports fp64 heat-bath and Metropolis throughput as a function of exchange shell count. One Monte Carlo sweep corresponds to one attempted update per spin. The cost of these routines is controlled not only by the number of interacting shells, but also by random-number generation, proposal construction, local field or energy evaluation and masked spin replacement. As a result, the shell-count dependence differs from the deterministic field-evaluation benchmarks.

Finally, Figure~3e benchmarks fp64 static minimization under fixed-iteration settings. Velocity projection, Barzilai--Borwein and L-BFGS are compared as implementation workloads rather than as globally tuned optimizer rankings. The goal is to quantify the cost of executing common minimization updates on large spin systems with the shared Hamiltonian backend. Convergence speed depends on optimizer parameters, stopping criteria and the energy landscape, whereas the fixed-iteration throughput reported here isolates the computational cost of each implementation.

Together, these benchmarks show that SpinX does not rely on a single universally optimal backend. Its performance comes from exposing multiple numerically consistent realizations of the same Hamiltonian and allowing the user to select the backend appropriate for the lattice regularity, interaction range, precision mode, hardware platform and simulation workflow. This multi-backend design is essential for covering short-range dynamics, extended frustrated exchange, dipolar fields, Monte Carlo sampling, static optimization and multi-image path calculations within one framework.    
 
\subsection*{Two competing annihilation channels of an exchange-stabilized hopfion}

We finally use SpinX to study annihilation pathways of an exchange-stabilized hopfion in a minimal frustrated-exchange model. The calculation is designed as a controlled atomistic test case rather than a material-specific prediction. The initial state is a metastable hopfion placed near the centre of a $100 \times 100 \times 100$ spin lattice, and the final state is the ferromagnetic state. Open boundary conditions are used, with a surrounding ferromagnetic background separating the hopfion from the sample boundary. This geometry helps distinguish annihilation of the hopfion tube from simple escape through the outer boundary.

Figure~4a shows two refined paths connecting the same initial and final states under the same Hamiltonian. The paths are constructed from string or NEB initializations and refined using GNEB. The blue path corresponds to a lateral-rupture channel, while the red path corresponds to an axial-collapse channel. The reaction coordinate is normalized independently for each path, and the energy is reported in units of the nearest-neighbour exchange $J_1$. The open stars mark the highest-energy images along the two paths. For the lateral-rupture channel, the high-energy region was further resolved by increasing the image density around the local maximum and applying local CI-GNEB refinement to better localize the saddle region. The magnified insets show the activation barriers of each channel, measured from the local metastable minimum on the hopfion side. The lower inset shows the saddle-energy difference $|\Delta E_{\mathrm{s}}| \approx 2.9\,J_1$ between the two channels.

The lateral-rupture pathway is shown in Figure~4b. Along this channel, the hopfion tube first develops a side-localized constriction. Near the highest-energy image, localized Bloch-point-like singular regions appear away from the central axis and rupture the toroidal texture from the side. The path then descends toward the ferromagnetic basin without an additional barrier comparable to the main saddle. This mechanism is distinct from boundary escape, since the rupture occurs on the hopfion tube itself rather than through motion of the entire texture out of the sample.

The axial-collapse pathway in Figure~4c provides a reference channel in the same model. In this case, the texture evolves more symmetrically along the path: the hopfion core shrinks along the axial direction, and Bloch-point-like singular regions appear near the central axis before the texture collapses toward the ferromagnetic state. This pathway resembles previously reported axial or core-collapse mechanisms of exchange-stabilized hopfions. In contrast, the lateral-rupture channel involves a side-localized rupture of the toroidal tube and leads to a distinct saddle morphology and activation barrier.

The coexistence of these two paths shows that hopfion annihilation in the same frustrated-exchange model is not exhausted by a single pathway. The energy landscape contains both an axial-collapse route and a lateral-rupture route, each involving a different topology-changing event. This result demonstrates the utility of large-scale atomistic path calculations in SpinX, where many-image transition-path workflows can be carried out on million-spin configurations and used to resolve secondary annihilation channels that are difficult to identify from relaxation or direct dynamics alone.

\section*{Methods}

\subsection*{Spin Hamiltonian and effective fields}

SpinX represents each atomic magnetic moment as
\[
\boldsymbol{\mu}_i
=
\mu_i\mu_{\mathrm B}\mathbf{S}_i ,
\]
where \(\mu_i\) is the moment magnitude in units of the Bohr magneton
\(\mu_{\mathrm B}\), and \(\mathbf{S}_i\) is a unit vector,
\(|\mathbf{S}_i|=1\). Throughout this work, spin Hamiltonian parameters are
stored internally in meV. The Bohr magneton is therefore written as
\[
\mu_{\mathrm B}^{\mathrm{meV/T}}
\approx
5.7884\times10^{-2}\,\mathrm{meV/T},
\]
when converting energy gradients to magnetic fields in tesla.

The total Hamiltonian is written as a sum of enabled contributions,
\[
\mathcal{H}
=
\mathcal{H}_{\mathrm{bil}}
+
\mathcal{H}_{\mathrm{Z}}
+
\mathcal{H}_{\mathrm{ani}}
+
\mathcal{H}_{\mathrm{dip}}
+
\mathcal{H}_{\mathrm{ho}},
\]
where \(\mathcal{H}_{\mathrm{bil}}\) contains pairwise bilinear interactions,
\(\mathcal{H}_{\mathrm{Z}}\) is the Zeeman coupling,
\(\mathcal{H}_{\mathrm{ani}}\) denotes onsite anisotropy,
\(\mathcal{H}_{\mathrm{dip}}\) is the dipolar interaction, and
\(\mathcal{H}_{\mathrm{ho}}\) collects supported higher-order terms. Terms that
are not enabled in a given calculation contribute zero\cite{Heisenberg1928ZPhys}.

For an unordered pair representation, the pairwise bilinear Hamiltonian is
\[
\mathcal{H}_{\mathrm{bil}}
=
-\sum_{i>j}
\sum_{\alpha,\beta}
J_{ij}^{\alpha\beta}
S_i^\alpha
S_j^\beta ,
\]
where \(J_{ij}^{\alpha\beta}\) is a \(3\times3\) interaction tensor in meV.
Equivalently, writing \(\mathsf{J}_{ij}\) for the matrix with elements
\(J_{ij}^{\alpha\beta}\),
\[
\mathcal{H}_{\mathrm{bil}}
=
-\sum_{i>j}
\mathbf{S}_i^{\mathsf T}
\mathsf{J}_{ij}
\mathbf{S}_j .
\]
In the implementation, unordered pair interactions are expanded into reciprocal
directed bonds, and the equivalent ordered-pair form is
\[
\mathcal{H}_{\mathrm{bil}}
=
-\frac{1}{2}
\sum_{i,j}
\sum_{\alpha,\beta}
J_{ij}^{\alpha\beta}
S_i^\alpha
S_j^\beta ,
\qquad
J_{ji}^{\beta\alpha}
=
J_{ij}^{\alpha\beta}.
\]
The factor \(1/2\) appears only in this ordered-pair representation and avoids
double counting of reciprocal directed bonds.

For scalar exchange and the Dzyaloshinskii-Moriya interaction, the tensor
action is defined by
\[
\sum_\beta
J_{ij}^{\alpha\beta}
S_j^\beta
=
J_{ij}S_j^\alpha
+
\left(
\mathbf{S}_j\times\mathbf{D}_{ij}
\right)^\alpha .
\]
Equivalently,
\[
\mathsf{J}_{ij}
=
J_{ij}\mathsf{I}
+
\begin{pmatrix}
0 & D_{ij}^z & -D_{ij}^y \\
-D_{ij}^z & 0 & D_{ij}^x \\
D_{ij}^y & -D_{ij}^x & 0
\end{pmatrix}.
\]
With this convention, the energy of an unordered interacting pair is
\[
\mathcal{H}_{ij}
=
-
J_{ij}\mathbf{S}_i\cdot\mathbf{S}_j
-
\mathbf{D}_{ij}\cdot
(\mathbf{S}_i\times\mathbf{S}_j),
\]
with \(\mathbf{D}_{ji}=-\mathbf{D}_{ij}\). Symmetric anisotropic exchange is
included through the symmetric part of \(J_{ij}^{\alpha\beta}\).

The Zeeman contribution is
\[
\mathcal{H}_{\mathrm{Z}}
=
-
\sum_i
\mu_i\mu_{\mathrm B}^{\mathrm{meV/T}}
\mathbf{B}^{\mathrm{ext}}_i
\cdot
\mathbf{S}_i .
\]
The onsite anisotropy used by SpinX has the form
\[
\mathcal{H}_{\mathrm{ani}}
=
-
\sum_i
\sum_\ell
K_{i\ell}
\left(
\mathbf{S}_i\cdot\hat{\mathbf n}_{i\ell}
\right)^2 ,
\]
where \(K_{i\ell}\) is stored in meV and \(\hat{\mathbf n}_{i\ell}\) is the
anisotropy axis. Additional supported terms include dipolar interactions and
selected higher-order interactions, including scalar biquadratic,
scalar-chirality and quadruplet terms in workflows where they are enabled.
Not all higher-order terms are available in every computational backend.

The effective field \cite{Antropov1996PRB}is defined in the atomistic spin-dynamics convention as
\[
\mathbf{B}_i^{\mathrm{eff}}
=
-
\frac{1}
{\mu_i\mu_{\mathrm B}^{\mathrm{meV/T}}}
\frac{\partial \mathcal{H}}
{\partial \mathbf{S}_i},
\]
where \(\mathbf{B}_i^{\mathrm{eff}}\) is in tesla when \(\mathcal H\) is in meV.
For the bilinear tensor term this gives
\[
B_{i,\alpha}^{\mathrm{bil}}
=
\frac{1}
{\mu_i\mu_{\mathrm B}^{\mathrm{meV/T}}}
\sum_{j,\beta}
J_{ij}^{\alpha\beta}
S_j^\beta .
\]
SpinX first evaluates the reduced bilinear field in meV,
\[
b_{i,\alpha}^{\mathrm{bil}}
=
\sum_{j,\beta}
J_{ij}^{\alpha\beta}
S_j^\beta ,
\]
and then converts it to a physical field through
\[
\mathbf{B}_i^{\mathrm{bil}}
=
\frac{\mathbf{b}_i^{\mathrm{bil}}}
{\mu_i\mu_{\mathrm B}^{\mathrm{meV/T}}}.
\]
The same effective-field convention is used by the dynamics, static
optimization, spectral-analysis and minimum-energy-path modules. Monte Carlo
sampling uses the same Hamiltonian parameters through local fields or local
energy differences, depending on the update scheme.

\subsection*{Backend representations}

For translationally invariant crystals, SpinX decomposes the magnetic lattice
into crystallographic sublattice channels. A spin on sublattice \(a\) in unit
cell \(\mathbf{R}\) is denoted \(\mathbf{S}_a(\mathbf{R})\). The reduced
bilinear field on sublattice \(a\) can be written as a multi-channel
convolution\cite{Cooley1965MathComp},
\[
b_a^\alpha(\mathbf{R})
=
\sum_b
\sum_{\boldsymbol{\delta}}
\sum_\beta
K_{ab}^{\alpha\beta}(\boldsymbol{\delta})
S_b^\beta(\mathbf{R}+\boldsymbol{\delta}),
\]
where \(K_{ab}^{\alpha\beta}(\boldsymbol{\delta})\) stores the interaction
tensor between sublattices \(a\) and \(b\) at displacement
\(\boldsymbol{\delta}\). Dense convolution, sparse convolution and FFT
convolution are different numerical realizations of this same expression.

The dense backend stores compact regular kernels and evaluates the reduced
field using convolution primitives. The sparse backend stores only non-zero
displacement entries and evaluates structured finite-range kernels without
forming a dense stencil. The FFT backend is used for fully periodic
translationally invariant interactions by transforming the spin field to
reciprocal space, multiplying by the precomputed kernel spectrum and
transforming back,
\[
\hat b_a^\alpha(\mathbf{q})
=
\sum_b
\sum_\beta
\hat K_{ab}^{\alpha\beta}(\mathbf{q})
\hat S_b^\beta(\mathbf{q}).
\]
Long-range dipolar fields are evaluated using the same reciprocal-space
convolution principle in workflows where the dipolar FFT path is enabled\cite{Berkov1993JMMM,GarciaCervera2003JCP}.

Finite regular supercells with open boundaries can still be handled by the
regular sublattice backends through the appropriate padding or truncation rule.
When translational regularity is broken by chemical disorder, structural
defects or explicitly site-resolved couplings, SpinX uses a generalized
pairlist backend. Each pairlist entry stores the source site, target site and
corresponding Hamiltonian parameters. The reduced bilinear field is evaluated as
\[
b_i^\alpha
=
\sum_{p:i_p=i}
\sum_\beta
J_p^{\alpha\beta}
S_{j_p}^\beta ,
\]
where \(p\) indexes pairlist entries, \(i_p\) is the target site and \(j_p\)
is the source neighbour associated with entry \(p\). This backend preserves the
same Hamiltonian interface while removing the requirement of translational
symmetry.

\subsection*{Atomistic spin dynamics}

Deterministic dynamics are integrated using the Landau-Lifshitz-Gilbert
equation\cite{Landau1935PhysZSU,Gilbert2004IEEE},
\[
\frac{d\mathbf{S}_i}{dt}
=
-\frac{\gamma}{1+\alpha_i^2}
\left[
\mathbf{S}_i \times \mathbf{B}_i^{\mathrm{eff}}
+
\alpha_i
\mathbf{S}_i \times
\left(
\mathbf{S}_i \times \mathbf{B}_i^{\mathrm{eff}}
\right)
\right],
\]
where \(\gamma\) is the gyromagnetic ratio and \(\alpha_i\) is the Gilbert
damping parameter. The main time integrator used in this work is a normalized
Heun scheme. Defining the right-hand side of the LLG equation as
\(\mathbf{f}_i(\mathbf{S})\), the predictor step is
\[
\tilde{\mathbf{S}}_i
=
\mathcal{N}
\left[
\mathbf{S}_i^n
+
\Delta t\,
\mathbf{f}_i(\mathbf{S}^n)
\right],
\]
followed by the corrected step
\[
\mathbf{S}_i^{n+1}
=
\mathcal{N}
\left[
\mathbf{S}_i^n
+
\frac{\Delta t}{2}
\left(
\mathbf{f}_i(\mathbf{S}^n)
+
\mathbf{f}_i(\tilde{\mathbf{S}})
\right)
\right].
\]
Here
\[
\mathcal{N}[\mathbf{x}]
=
\frac{\mathbf{x}}{|\mathbf{x}|}
\]
denotes projection back to the unit sphere.

\subsection*{Stochastic Landau-Lifshitz-Gilbert simulations}

For stochastic LLG simulations, a Gaussian thermal field is added to the
effective field\cite{Brown1963PhysRev,Evans2014JPCM},
\[
\mathbf{B}_i^{\mathrm{eff}}
\rightarrow
\mathbf{B}_i^{\mathrm{eff}}
+
\mathbf{B}_i^{\mathrm{th}} .
\]
In discrete time, the thermal field is written as
\[
\mathbf{B}_i^{\mathrm{th},n}
=
\sigma_i
\boldsymbol{\eta}_i^n ,
\]
where the Cartesian components of \(\boldsymbol{\eta}_i^n\) are independent
standard normal random variables. For the LLG convention used here,
\[
\sigma_i
=
\sqrt{
\frac{2\alpha_i k_{\mathrm B}T}
{|\gamma|\mu_i\mu_{\mathrm B}^{\mathrm{meV/T}}\Delta t}
}.
\]
Here \(k_{\mathrm B}T\) is evaluated in meV. This noise amplitude corresponds
to the LLG convention used in the equation above. Random fields are generated
reproducibly from explicit JAX random keys. In mixed-precision mode,
effective-field evaluation is performed in fp32, while the spin state and time
integration are kept in fp64.

\subsection*{Monte Carlo sampling}

SpinX implements heat-bath and Metropolis Monte Carlo updates for supported
spin Hamiltonians\cite{Metropolis1953JCP,Creutz1987PRD}. The heat-bath update is used when the local energy is linear
in the updated spin, as in bilinear exchange models with optional Zeeman terms.
The Monte Carlo benchmarks in this work use this class of Hamiltonians. In a
local update, the local field acting on site \(i\) is held fixed while a new
spin direction is sampled or proposed. For the heat-bath update, the spin is
sampled from
\[
P(\mathbf{S}_i)
\propto
\exp
\left[
\beta
\mathbf{b}_i^{\mathrm{loc}}
\cdot
\mathbf{S}_i
\right],
\]
where \(\mathbf{b}_i^{\mathrm{loc}}\) is the reduced local field in meV and
\(\beta=(k_{\mathrm B}T)^{-1}\), with \(k_{\mathrm B}T\) in meV. Equivalently,
using the physical local field \(\mathbf{B}_i^{\mathrm{loc}}\),
\[
\mathbf{b}_i^{\mathrm{loc}}
=
\mu_i\mu_{\mathrm B}^{\mathrm{meV/T}}
\mathbf{B}_i^{\mathrm{loc}} .
\]
In the Metropolis update, a trial spin \(\mathbf{S}'_i\) is accepted with
probability
\[
P_{\mathrm{acc}}
=
\min
\left[
1,
\exp(-\beta \Delta\mathcal{H}_i)
\right],
\]
where \(\Delta\mathcal{H}_i\) is the local energy change in meV associated with
the trial move.

For compatible regular lattices, SpinX uses colour-based site-major updates so
that spins updated simultaneously do not interact within the selected
interaction range. The number of colours is chosen according to the neighbour
range of the active Hamiltonian, enabling parallel updates while preserving the
local dependency structure of the Monte Carlo move.

\subsection*{Thermodynamic and dynamical observables}

SpinX stores the total magnetic moment as
\[
\mathbf{M}_{\mathrm{tot}}
=
\sum_i
\mu_i\mathbf{S}_i ,
\]
in units of \(\mu_{\mathrm B}\). For equal magnetic moments, the plotted
reduced magnetization is
\[
\mathbf{m}
=
\frac{1}{N}
\sum_i
\mathbf{S}_i,
\qquad
M
=
|\mathbf{m}|.
\]
For the temperature scans shown in Figure~2, the plotted magnetization,
susceptibility and specific heat are scaled for visual comparison.

The Binder cumulant is evaluated from the sampled magnetization magnitude as
\[
U_4
=
1
-
\frac{\langle M^4\rangle}
{3\langle M^2\rangle^2}.
\]
For the plotted finite-temperature curves, SpinX uses the following
fluctuation-based estimators before rescaling for visual comparison,
\[
\chi_{\mathrm{rep}}
=
\frac{N}{T}
\left(
\langle M^2\rangle
-
\langle M\rangle^2
\right),
\]
and
\[
C_{v,\mathrm{rep}}
=
\frac{N}{T^2}
\left(
\langle e^2\rangle
-
\langle e\rangle^2
\right),
\qquad
e=\frac{\mathcal{H}}{N}.
\]
These reported quantities are used consistently for the scaled curves in
Figure~2.

For dynamical spectra, the time average is first subtracted from each spin
component,
\[
\delta S_{i\alpha}(t)
=
S_{i\alpha}(t)
-
\langle S_{i\alpha}\rangle_t .
\]
The spatial Fourier amplitude is then computed as
\[
M_\alpha(\mathbf{q},t)
=
\frac{1}{\sqrt{N}}
\sum_i
\mu_i
\delta S_{i\alpha}(t)
\exp(-i\mathbf{q}\cdot\mathbf{r}_i).
\]
A Hann window is applied to \(M_\alpha(\mathbf{q},t)\) along \(t\) before the
time Fourier transform to reduce spectral leakage. The spectral tensor is
formed as
\[
S_{\alpha\beta}(\mathbf{q},\omega)
=
M_\alpha^*(\mathbf{q},\omega)
M_\beta(\mathbf{q},\omega).
\]
Overall normalization factors associated with the time window, window function
and trajectory length are omitted because the spectra in Figure~2 are plotted in
arbitrary units. For the ferromagnetic spectra shown in Figure~2, the equilibrium
magnetization is along the \(z\) axis and the plotted transverse intensity is
\[
S_{\perp}(\mathbf{q},\omega)
=
S_{xx}(\mathbf{q},\omega)
+
S_{yy}(\mathbf{q},\omega).
\]
More generally, SpinX can also form a \(q\)-transverse projection using
\(P_{\alpha\beta}=\delta_{\alpha\beta}-\hat q_\alpha\hat q_\beta\). The
numerical Fourier transform returns ordinary frequencies \(f\), and the plotted
energy is \(E=hf=\hbar\omega\).

\subsection*{Static optimization and minimum-energy paths}

Static minimization is performed on the product manifold of unit spin spheres\cite{Absil2008Optimization}.
Given the Euclidean gradient
\[
\nabla_i\mathcal{H}
=
\frac{\partial \mathcal{H}}
{\partial \mathbf{S}_i},
\]
the corresponding tangent gradient is
\[
\mathbf{G}_i
=
\nabla_i\mathcal{H}
-
(\nabla_i\mathcal{H}\cdot\mathbf{S}_i)\mathbf{S}_i .
\]
Optimization algorithms such as velocity projection, Barzilai-Borwein,
L-BFGS and Adam update tangent directions and retract the spin back to the unit
sphere\cite{Bitzek2006PRL,Barzilai1988IMA,Liu1989MathProg,Kingma2015Adam}. SpinX supports both normalized and exponential-map retractions. The
normalized retraction is
\[
\mathcal{R}^{\mathrm{norm}}_{\mathbf S_i}(\mathbf u_i)
=
\frac{\mathbf S_i+\mathbf u_i}
{|\mathbf S_i+\mathbf u_i|},
\]
whereas the exponential-map retraction is
\[
\mathcal{R}^{\mathrm{exp}}_{\mathbf S_i}(\mathbf u_i)
=
\cos|\mathbf u_i|\mathbf S_i
+
\sin|\mathbf u_i|
\frac{\mathbf u_i}{|\mathbf u_i|},
\qquad
\mathbf u_i\cdot\mathbf S_i=0 .
\]
For \(|\mathbf u_i|=0\), the exponential-map expression is evaluated in its
limiting form. The minimum-energy-path calculations in this work use the
exponential-map retraction unless otherwise stated.

Minimum-energy paths are computed using string and geodesic nudged elastic band
methods\cite{E2002PRB,Henkelman2000JCP,Bessarab2015CPC}. An image \(n\) is a full spin configuration
\[
\mathbf{Y}_n
=
\{\mathbf{S}_i^{(n)}\}_{i=1}^{N}.
\]
Distances between images are evaluated using the geodesic metric on the spin
manifold,
\[
d(\mathbf{Y}_n,\mathbf{Y}_m)
=
\left[
\sum_i
\arccos^2
\left(
\mathbf{S}_i^{(n)}\cdot\mathbf{S}_i^{(m)}
\right)
\right]^{1/2}.
\]

In the string method, each image is first displaced by the projected physical
force on the spin manifold. The path is then reparameterized along the
geodesic reaction coordinate so that the image distribution follows the chosen
arc-length spacing. This separates the relaxation of the path shape from the
maintenance of image spacing and avoids introducing an explicit spring force.

For image \(n\), the GNEB physical force is the negative energy gradient
projected onto the tangent space of the spin manifold. The GNEB force separates
this physical force into a component perpendicular to the path tangent and a
spring force parallel to the path,
\[
\mathbf{F}_n^{\mathrm{GNEB}}
=
\mathbf{F}_{n,\perp}^{\mathrm{phys}}
+
\mathbf{F}_{n,\parallel}^{\mathrm{spring}}.
\]
For climbing-image refinement, the highest-energy image is driven toward the
saddle region by reversing the tangential component of the physical force,
\[
\mathbf{F}_n^{\mathrm{CI}}
=
\mathbf{F}_n^{\mathrm{phys}}
-
2
\left(
\mathbf{F}_n^{\mathrm{phys}}
\cdot
\hat{\boldsymbol{\tau}}_n
\right)
\hat{\boldsymbol{\tau}}_n .
\]
The vector \(\hat{\boldsymbol{\tau}}_n\) is the normalized path tangent in image
space, and all forces are projected onto the spin tangent space. Endpoints can
be pinned, and image redistribution is applied when needed to maintain path
resolution.

\subsection*{Performance measurements}

Performance benchmarks report throughput in spin sites processed per second.
For effective-field benchmarks, one spin-site operation corresponds to one
evaluated effective field for one spin. For Heun dynamics, one spin-site
operation corresponds to one full Heun time step for one spin, including the
predictor-corrector update. For Monte Carlo, throughput is reported as
attempted spin updates per second. For static optimization, throughput is
reported as spin-site updates per fixed optimization iteration.

All timings exclude JIT compilation and include device synchronization after
execution. AMD MI250X results are reported per Graphics Compute Die (GCD),
whereas NVIDIA GH200 results are reported per GPU. The hardware and software
environment used for the benchmarks is summarized in Table~\ref{tab:hardware}.
Benchmark system sizes, precision modes, backend choices and workload 
definitions are specified in the corresponding figure captions.

\begin{table}[h]
\centering
\caption{Hardware and software environment for performance benchmarks.}
\begin{tabular}{lll}
\hline
Platform & Accelerator & Software stack \\
\hline
NVIDIA & GH200 GPU & JAX 0.9.2, CUDA 12 \\
AMD & MI250X GCD & JAX 0.9.2, ROCm 7\\
\hline
\label{tab:hardware}
\end{tabular}
\end{table}

\subsection*{Hopfion calculation}

The exchange-stabilized hopfion in Figure~4 is computed using a minimal
four-shell frustrated-exchange model on a simple-cubic lattice\cite{rybakov2022magnetic}. The model is
used as a controlled atomistic realization of a frustrated-exchange Hamiltonian
that stabilizes Hopf solitons, rather than as a material-specific parameter
set. The Hamiltonian used for this calculation is
\[
\mathcal{H}
=
-\sum_{i>j}
J_{ij}
\mathbf{S}_i\cdot\mathbf{S}_j ,
\]
with non-zero couplings on the first four neighbour shells. We use
\[
J_1 \simeq 12.4~\mathrm{meV}
\]
as a representative energy scale, and
\[
\frac{J_2}{J_1}=0.188,\qquad
\frac{J_3}{J_1}=-0.274,\qquad
\frac{J_4}{J_1}=-0.161 .
\]
All energies in Figure~4 are reported in units of \(J_1\).

The initial hopfion configuration with Hopf index \(Q_{\mathrm H}=1\) is
prepared from a toroidal Hopf-map ansatz,
\[
\mathbf{S}(\mathbf{r})
=
\left(
\cos\Phi(\mathbf{r})\sin\Theta(\mathbf{r}),
\;
\sin\Phi(\mathbf{r})\sin\Theta(\mathbf{r}),
\;
\cos\Theta(\mathbf{r})
\right),
\]
where \(\Theta(\mathbf{r})\) and \(\Phi(\mathbf{r})\) encode the linked-fibre
structure of the Hopf map. The ansatz is placed near the centre of a
\(100\times100\times100\) simple-cubic lattice with open boundary conditions
and a surrounding ferromagnetic region large enough to reduce spurious boundary
effects on the annihilation mechanism. The ansatz is relaxed using static
minimization with the velocity-projection method to obtain the metastable
hopfion endpoint. The final endpoint is the uniform ferromagnetic state under
the same Hamiltonian and boundary conditions.

Transition paths between the relaxed hopfion and the ferromagnetic state are
generated from multiple trial paths and refined using string and geodesic
nudged elastic band calculations. Two reproducible annihilation channels are
retained for analysis: an axial-collapse channel and a lateral-rupture channel.
The paths shown in Figure~4 are assembled from GNEB-refined segments with pinned
endpoints. For the lateral-rupture channel, the image density near the
highest-energy region is increased and a local CI-GNEB refinement is applied to
the saddle region. In this local refinement, the maximum perpendicular force on
the climbing image is reduced below \(5\times10^{-6}\,J_1\) per spin. Outside
the locally refined CI region, the maximum perpendicular force along the path is
kept below \(\sim10^{-3}\,J_1\) per spin. Here forces are measured as energy
gradients with respect to dimensionless spin coordinates.

Bloch-point-like regions are identified from localized peaks in a discretized
topological-charge diagnostic on the magnetic lattice. The spin textures in
Figure~4 are rendered as streamtubes traced on the \(m_z=0\) isosurface, with
colours representing the local in-plane spin phase. These markers are used to
visualize topology-changing regions along the transition path\cite{Berg1981NuclPhysB,Kuchkin2025PRR}.

\section{Code and data availability}

The source code of SpinX v0.1.0, corresponding to the version used for the calculations in this work, is publicly available at: https://github.com/QichenXu-Research/SpinX. This repository is released as an initial review snapshot and will continue to evolve during and after the review process. The repository includes the core SpinX package, command-line interface, validation tests and minimal executable examples.

\section{acknowledgments}
The authors thank Filipp N. Rybakov (Uppsala University), Olle Eriksson (Uppsala University), Manuel Pereiro(Uppsala University), Pavel Bessarab (Linnaeus University) and Liuzhen Yang for many fruitful discussions. We also thank Johan Hellsvik (KTH, Dardel), Olle Eriksson (CSC, LUMI) for support with GPU resources. The authors used AI-assisted tools to improve the language of the manuscript.

Financial support from the Swedish Research Council (Vetenskapsrådet, VR; Grant No. 2024-04986), and the Knut and Alice Wallenberg Foundation (KAW; Grant No. 2022.0108), is acknowledged. 
The Wallenberg Initiative Materials Science for Sustainability (WISE) funded by the Knut and Alice Wallenberg Foundation and The Swedish e-Science Research Centre (SeRC) are also acknowledged.
The computations/data handling were enabled by resources provided by the National Academic Infrastructure for Supercomputing in Sweden (NAISS), partially funded by the Swedish Research Council through grant agreement no. 2022-06725.

\section{Competing Interests}
All authors declare that they have no conflicts of interest.
\bibliography{SA}
\end{document}